\documentclass[twoside,a4paper,11pt]{torus2012}
\usepackage{graphicx}
\usepackage{hyperref}
\usepackage{movie15}
\def\apj{\mbox{ApJ}}
\def\apjl{\mbox{ApJL}}
\def\apjs{\mbox{ApJS}}
\def\aaps{\mbox{A\&AS}}
\def\mnras{\mbox{MNRAS}}

\def\araa{\mbox{ARA\&A}}
\def\pasp{\mbox{PASP}}
\def\nat{\mbox{Nature}}
\def\aap{\mbox{A\&A}}

\def\mum{$\mu$m }
\def\mums{$\mu$m}

\begin{document}
\pagenumbering{arabic}
\pagestyle{myheadings}
\thispagestyle{empty}
{\flushleft\includegraphics[width=\textwidth,bb=58 650 590 680]{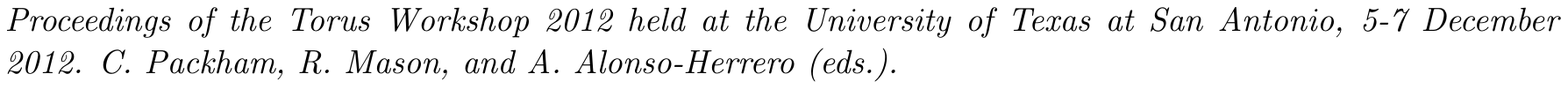}}
\vspace*{0.2cm}
\begin{flushleft}
{\bf {\LARGE
%
Modeling IR SED of AGN with \textit{Spitzer} and \textit{Herschel} data
%
}\\
\vspace*{1cm}
%
Feltre A.$^{1,2}$
%
}\\
\vspace*{0.5cm}
%
$^{1}$ESO, Karl-Schwarzschild-Str. 2, 85748 Garching bei M\"unchen, Germany\\
$^{2}$Dipartimento di Fisica e Astronomia, Vicolo Osservatorio 2, I-35122 Padova, Italy\\
%
\end{flushleft}
%
\markboth{
Modelling IR SED of AGN
}{ 
%
Feltre et al.
%
}
\thispagestyle{empty}
\vspace*{0.4cm}
\begin{minipage}[l]{0.09\textwidth}
\ 
\end{minipage}
\begin{minipage}[r]{0.9\textwidth}
\vspace{1cm}
\section*{Abstract}{\small
%

One of the remaining open issues in the context of the analysis of Active Galactic Nuclei (AGN) is the evidence that nuclear gravitational accretion is often accompanied by a concurrent starburst (SB) activity. What is, in this picture, the role played by the obscuring dust around the nucleus and what do the state of the art AGN torus models have to say? Can the IR data provided by \textit{Spitzer} and \textit{Herschel} help us in extensively investigate both phenomena and, if so, how and with what limitations? In this paper we present our contribution to the efforts of answering these questions. We show some of the main results coming from a comparative study of various AGN SED modeling approaches, focusing mostly on the much-debated issue about the morphology of the dust distribution in the toroidal structure surrounding the AGN. We found that the properties of dust in AGN as measured by matching observations (be it broad band IR photometry or IR spectra) with models, strongly depend on the choice of the dust distribution. Then, we present the spectral energy distribution (SED) fitting procedure we developed, making make the best use of \textit{Spitzer} and \textit{Herschel} SPIRE mid- and far-IR observations, to dig into the role played by the possible presence of an AGN on the host galaxy's properties. 
%
\normalsize}
\end{minipage}
%
%
%
\section{Introduction \label{intro}}

The central black hole of Active Galactic Nuclei (AGN) is belived to be surrounded by a dusty structure, commonly referred to us as ``AGN dusty torus'', which, according to the unified scheme for AGN \cite{antonucci93} plays a crucial role in explaining the difference between type 1 and type 2 AGN as an orientation effect \cite{rees69}. The intersection of the line of sight with the obscuring dust can give rise to the observed lack of broad lines in the optical spectra, typical of type 2 objects, and to a number of differences in the observed SED of AGN at almost all wavelengths.

The various radiative transfer models of AGN tori that have been developed in the last decades can be divided mainly into two categories: those in which the dust is uniformly distributed in an axisymmetric geometry around the central supermassive black hole, customarily referred to as ``smooth'' models"\cite{pier92,granato94,stenholm94,efstathiou95,manske98,vanbemmel03,schartmann05,fritz06}, the first that has been developed because more simple to compute and those in which the dust is distributed in clumps or clouds surrounding the central source, and hence referred to as ``clumpy models" \cite{nenkova02,dullemond05,honig06,schartmann08}. In the first part of this paper we report the main results coming from a systematic comparison (see Feltre et al. 2012 for more details \cite{feltre12}) between two of the most widely used models in the literature, each one representative of one of the two classes.

In order to get a more complete view of the phenomena we also perform a multiwavelength study developing an {\itshape ad hoc} SED fitting procedure aimed at reproducing the SEDs of the AGN from the optical/UV to the far-IR. Thanks to its multi-component approach this tool turned out to be very powerful for the investigation of the interplay between the AGN and the host galaxy. The methodology and the first preliminary results are presented in the second part of the paper.

\section{AGN Torus models comparison}\label{sec:models}

In order to investigate the possibility to use the SED fitting technique as a diagnostic for the dust distribution we carried out a systematic comparison between two of the most popular models in the literature, namely: smooth models of Fritz et al. 2006 \cite{fritz06}, hereafter F06, and clumpy models of Nenkova et al. 2008 \cite{nenkova02,nenkova08a,nenkova08b}, hereafter N08. We first conducted an automatic comparison (Sec. \S \ref{sec:modelmodel}) investigating the main properties of the model SEDs and, in a second step, we compared the models with observations considering mid-IR spectra information and photometric data (Sec. \S \ref{sec:modelobs}). 


\subsection{Grid selection}

\begin{figure}
\center
\includegraphics[width=6cm,angle=270]{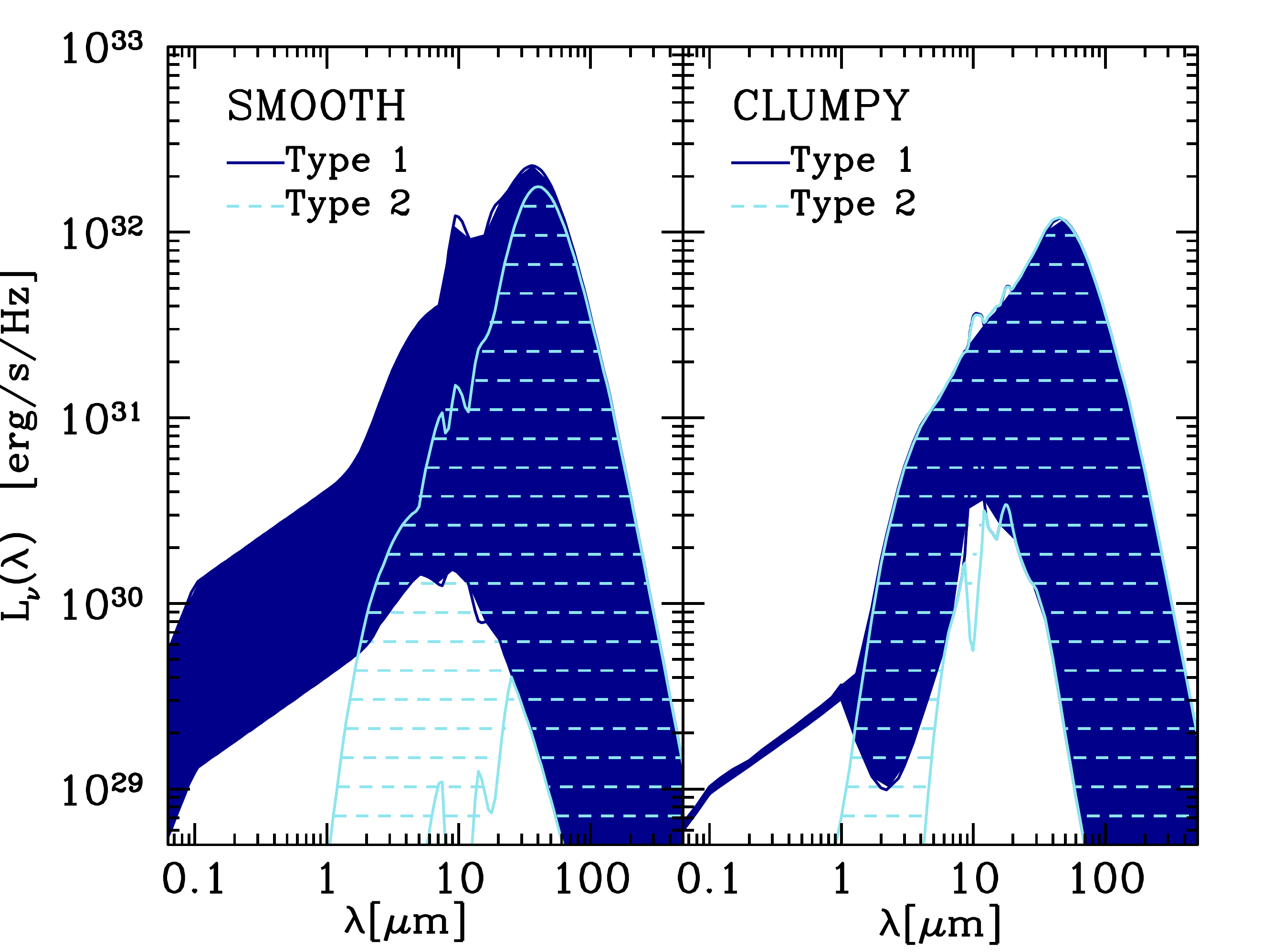} 
\caption{\label{fig1} The range of SEDs covered by the smooth (left) and clumpy (right) dust configurations in the restricted parameter grids. The coverage is shown for both type 1 (filled regions) and type 2 (dashed) inclinations \cite{feltre12}.
}
\end{figure}

The comparison was carried out considering two grids of the aforementioned models, F06 and N08, selected trying to match the values of the model parameters, such as the outer-inner radius ratio, the optical depth along the line of sight and the torus aperture. 
Furthermore, in the model-to-model comparison we only considered  two extreme inclinations, corresponding to type 1 and type 2 objects. Fig. \ref{fig1} shows region occupied by the two model grids for type 1 objects and type 2 views. 

\subsection{Model-to-model comparison}\label{sec:modelmodel}

\begin{figure}
\centerline{
\includegraphics[angle=270,width=6cm]{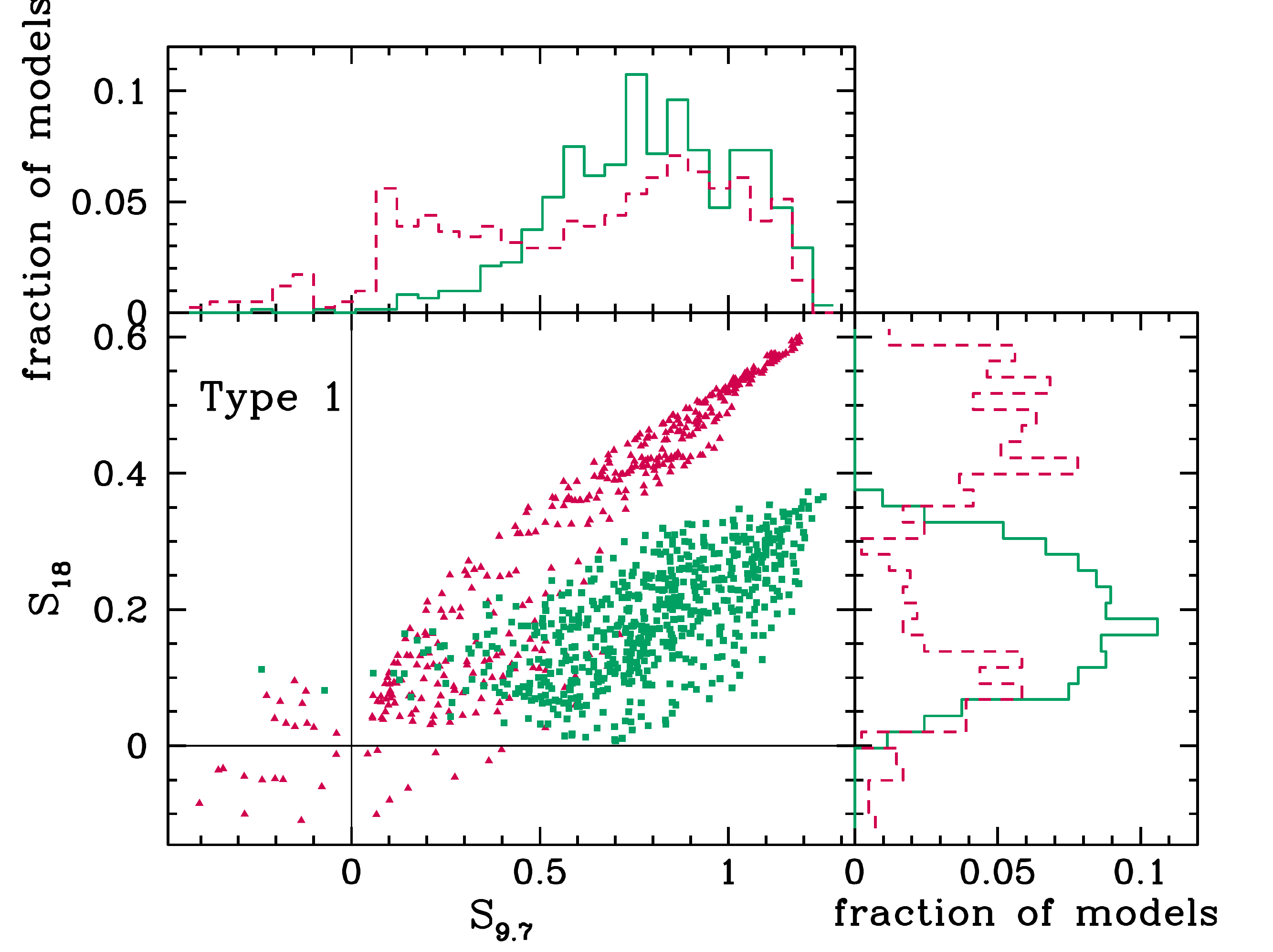}\hspace{1cm}
\includegraphics[angle=270,width=6cm]{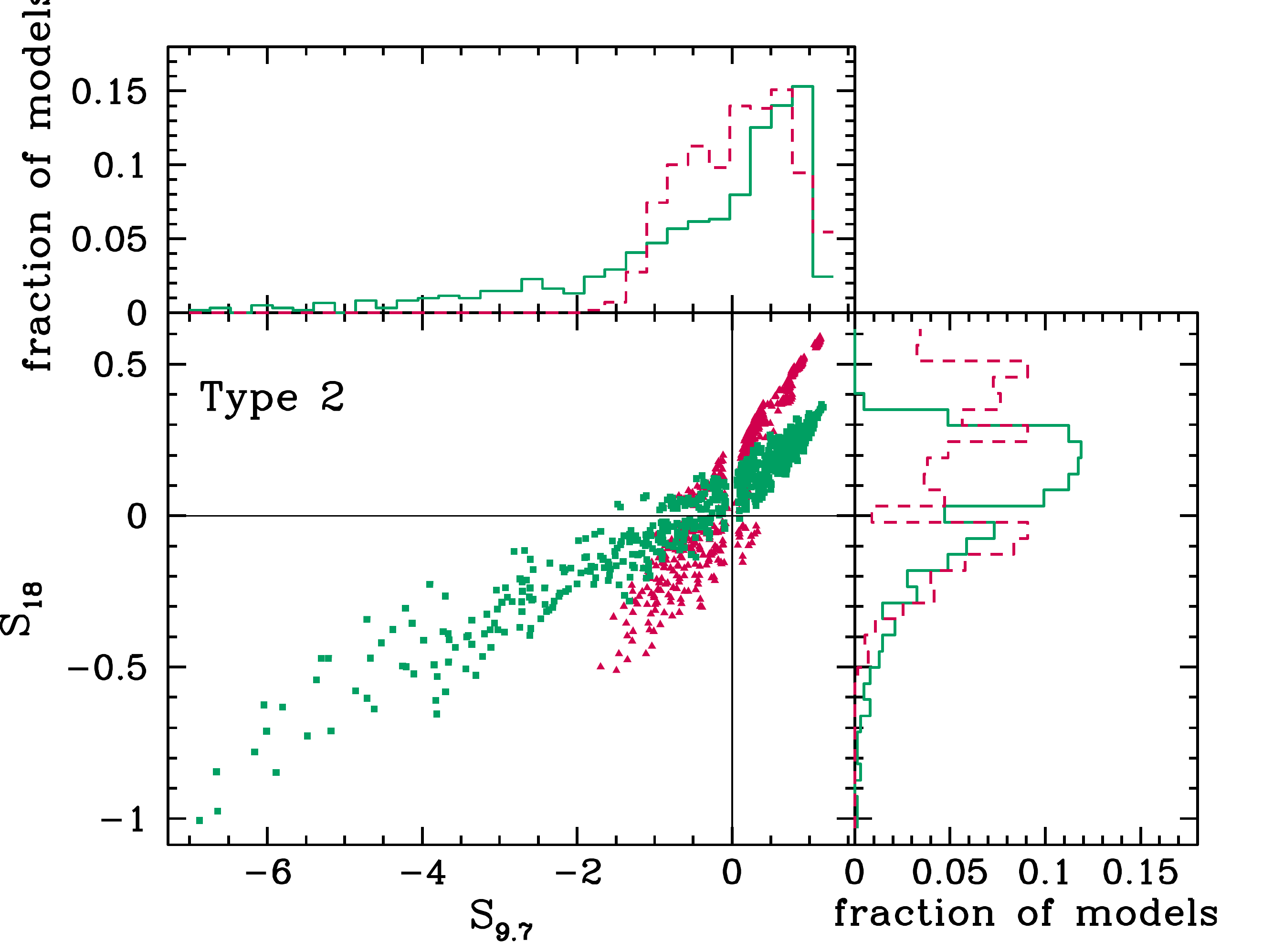}}
\caption{$S_{18}$ versus $S_{9.7}$ for smooth (green squares) and clumpy (red triangles) models,  and $S_{9.7}$ and $S_{18}$ distributions for type 1 (left) and type 2 (right) views.\cite{feltre12}}
\label{fig2}
\end{figure}

In Feltre et al. 2012 \cite{feltre12} we investigated different properties of the IR SED of the models, that is the width and the peak wavelength of the IR SED, the strength of the silicate features at 9.7 and 18 \mum \footnote{Defined as the logarithm of the ratio between the flux $F$ measured within the line profile over the continuum flux $F_{c}$ at such wavelength, i.e.
\begin{equation}
S=\mbox{ln}\left(F(\lambda_m)/F_c(\lambda_m)\right)
\label{eq:S}
\end{equation}
where $\lambda_m$ is the wavelength at which the feature's strength is an extremum with a value in the interval between 8.5 and 11.5 \mum for $S_{9.7}$ and  between 17.0 and 19.5 \mum for $S_{18}$.[for the computation of $F_{c}$ see Sirocky et al. 2008 \cite{sirocky08}. F06 and N08 models consider different absorption coefficients that peak at different wavelengths, but for simplicity we call the strength of the silicate feature around 9.7 \mum $S_{9.7}$, irrespective of $\lambda_m$.}, the spectral index in the mid-IR $\alpha_{IR}$\footnote{Defined as

\begin{equation}
	 \alpha_{IR}=\frac{\log_{10}(F_{4.5})-\log_{10}(F_{3.6})}{\log_{10}(\lambda_{4.5})-\log_{10}(\lambda_{3.6})}.
\end{equation}} and the luminosity at 12.3 \mums. Our findings can be summarised as follows:

\noindent - $S_{9.7}$ and $S_{18}$ of both F06 and N08 models are shown in the feature-feature diagram in Fig. \ref{fig2} for both type 1 (left) and type 2 (right) views. In particular, for type 1 views F06 and N08 models occupy different region of the diagram. This difference arises from the fact that the two models use different silicate absorption coefficients: Laor \& Draine 1993 \cite{laor93} and Ossenkopf 1992 \cite{ossenkopf92} for F06 and N08, respectively;

\noindent - the distribution of the width of the IR SEDs, $W_{IR}$, is narrower for N08 models, for both type 1 and 2 views, despite the presence of a range of values common with the $W_{IR}$ distribution of F06 models; 

\noindent - the distribution of $\alpha_{IR}$, especially for type 1 views, is very different between F06 and N08 models and the common range of values is very narrow. This difference is mostly likely due to the two different power laws used to model the primary source and lack of a very hot component in the clumpy models (see e.g. \cite{deo11}).


\begin{figure}
\centerline{
\includegraphics[angle=270,width=6cm]{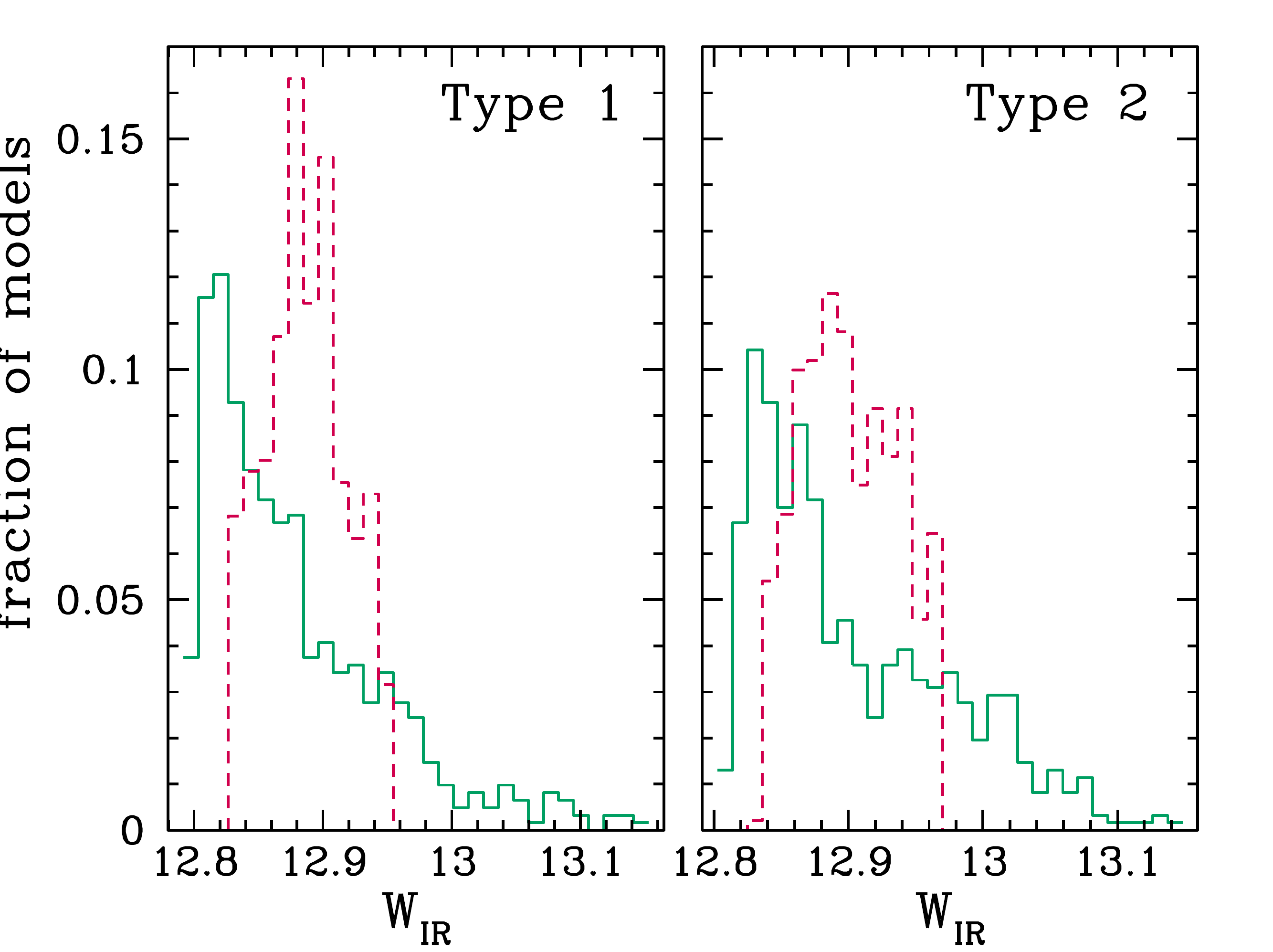}\hspace{1cm}
\includegraphics[angle=270,width=6cm]{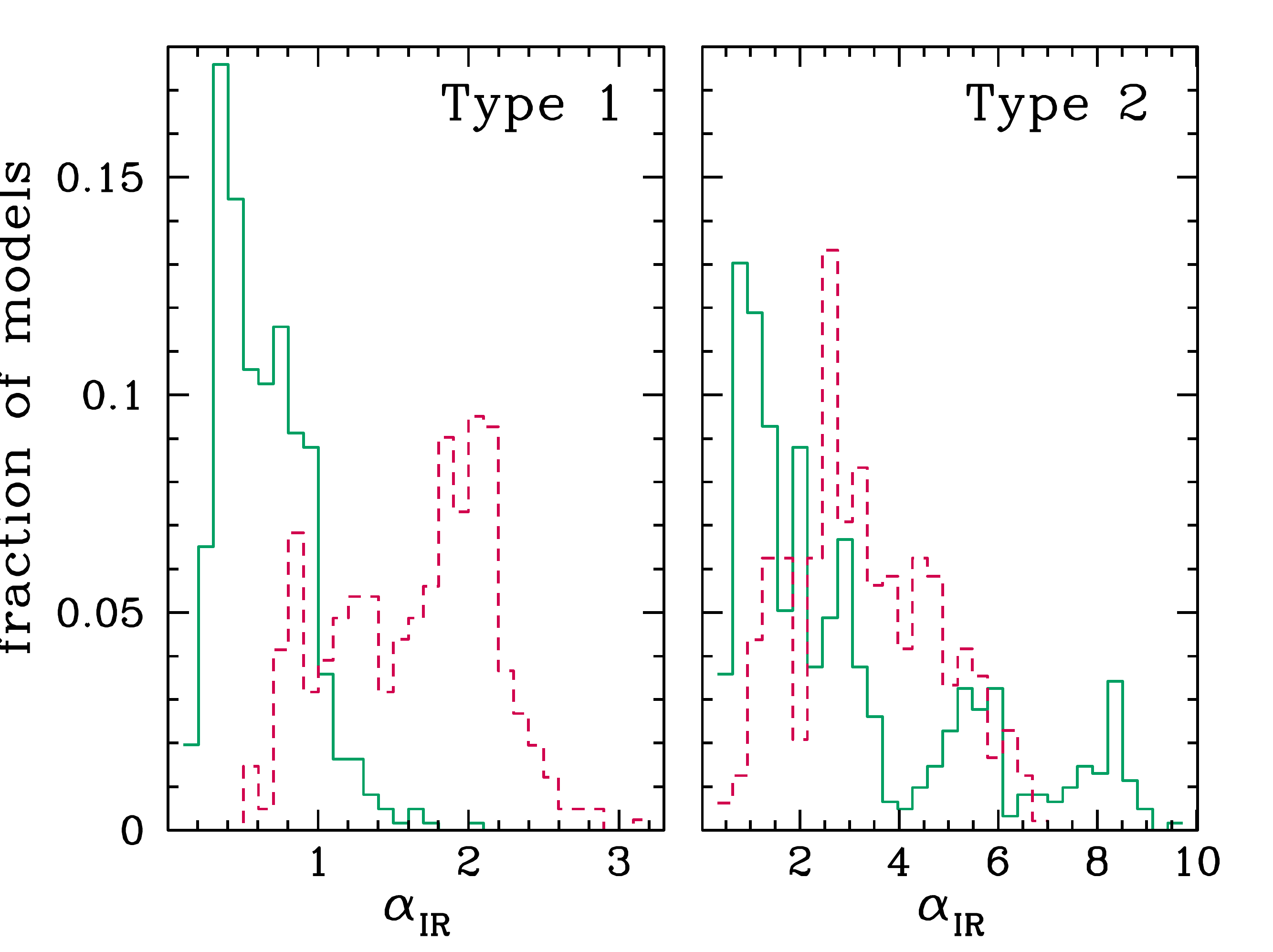}}
\caption{W$_{IR}$ and  $\alpha_{IR}$, for type 1 (left column) and type 2 (right column) views, for smooth (continuous green lines) and clumpy (dashed red lines) models.\cite{feltre12}}
\label{fig4}
\end{figure}

\begin{figure}
\centerline{
\includegraphics[width=6cm]{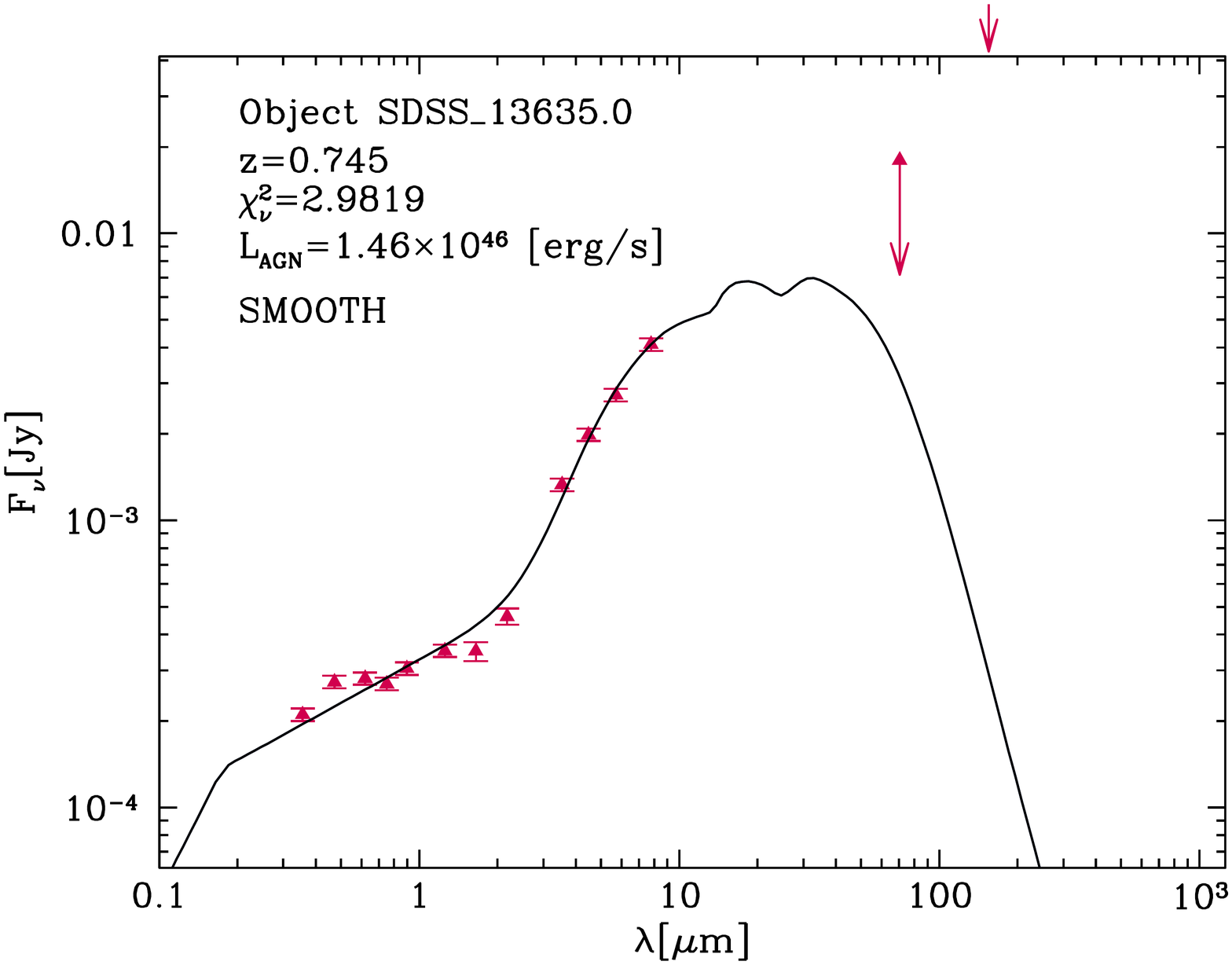}
\includegraphics[width=6cm]{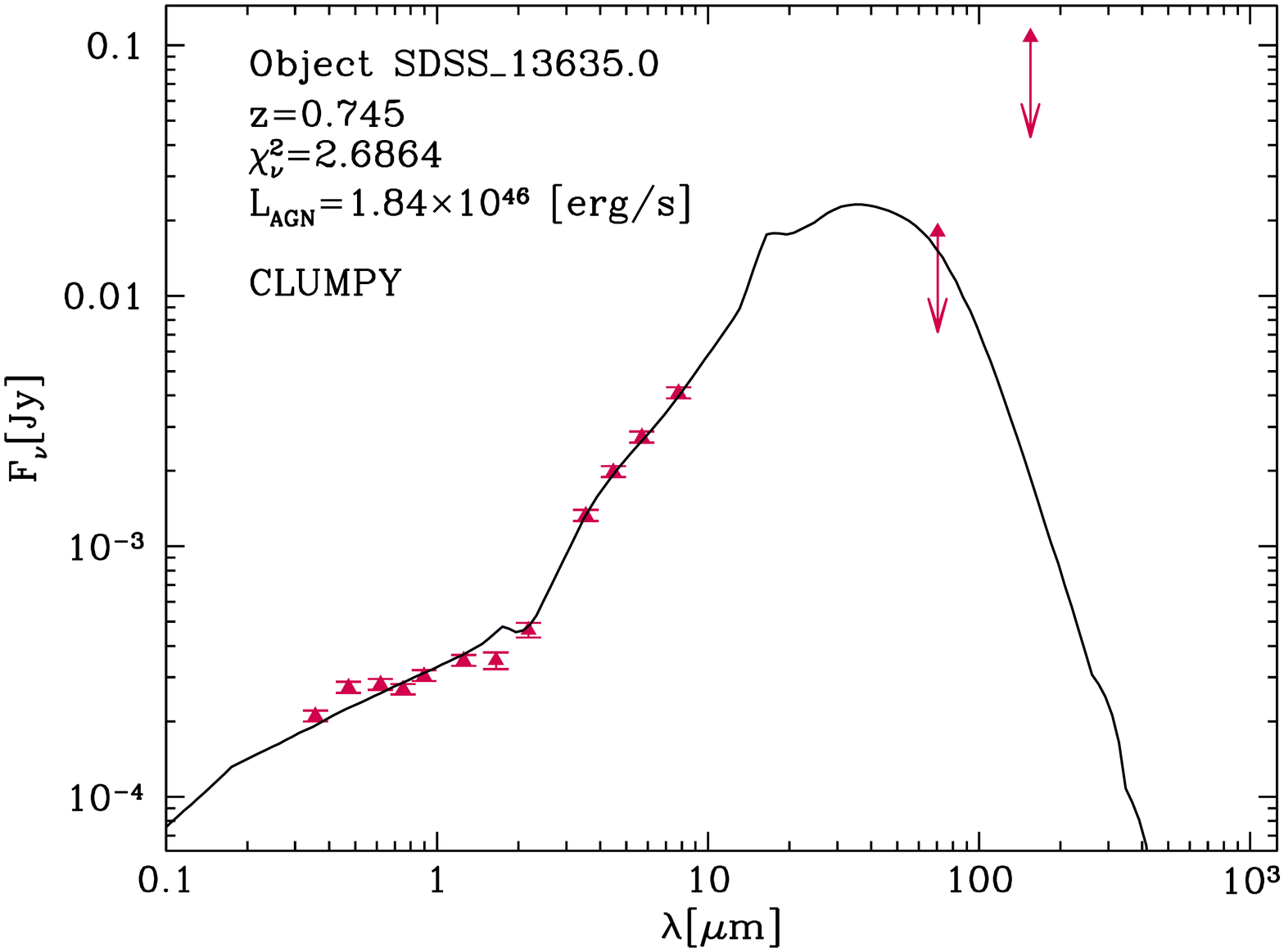}}
\caption{Object SDSS\_13635.0 (RA = 10:34:13.89(hms), Dec = +58:52:52.8(dms)) is reproduced by a single component, namely that of a type 1 AGN F06 model (left) and N08 model (right).\cite{feltre12}}
\label{fig6}
\end{figure}

\subsection{Comparison with observations}\label{sec:modelobs}

\begin{figure}
\centerline{
\includegraphics[width=7.5cm]{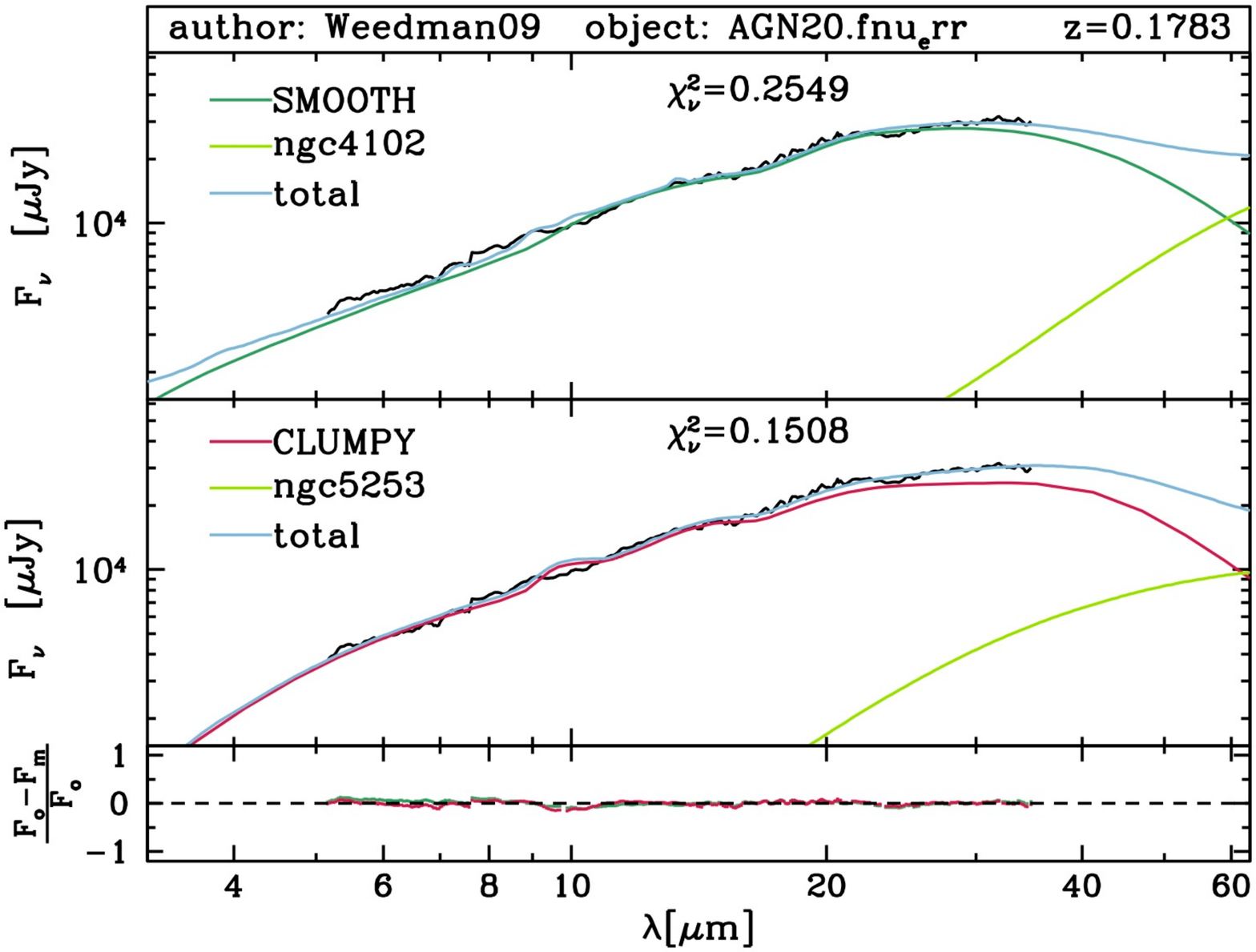}}
\caption{Example of the SED fitting of the IRS spectrum of AGN24 \cite{weedman09} for F06 models (top panel - dark green line) and N08 models (bottom panel - red line).\cite{feltre12}}
\label{fig8}
\end{figure}

The second step is to compare the models with observations. We firstly consider a sample of 278 spectroscopically confirmed SDSS type 1 quasars with redshift spanning $0.06<z<5.2$ with photometry coming from SDSS, 2MASS (whenever available) and SWIRE (Spitzer Wide-area InfraRed Extragalactic Survey \cite{lonsdale04}). This sample has been presented in Hatziminaoglou et al. 2008 \cite{hatzimi08}. We also consider a sample of 160 objects with mid-IR spectra which was a part of a collection of IRS spectra available in the literature (covering the wavelength range from $\sim 5$ to $\sim 40$ $\mu$m) presented in Hern\'an-Caballero \& Hatziminaoglou 2011 \cite{hernan11}. As can be seen in the examples reported in Figs. \ref{fig6} and \ref{fig8} both "smooth" and "clumpy" models provide equally good fits when photometric data or IRS spectra are considered.

\section{Multiwavelength emission of AGN}\label{sec:fit}

The availability of a large quantity of photometric data points available requires a customised and versatile tool to investigate the multi-wavelength properties of extragalactic sources. We present here a multi-component multi-wavelength SED fitting procedure accounting for the different emission mechanisms acting from the UV to the far-IR.

\subsection{SED fitting procedure}

\begin{figure}
\centering
\includegraphics[width=6cm]{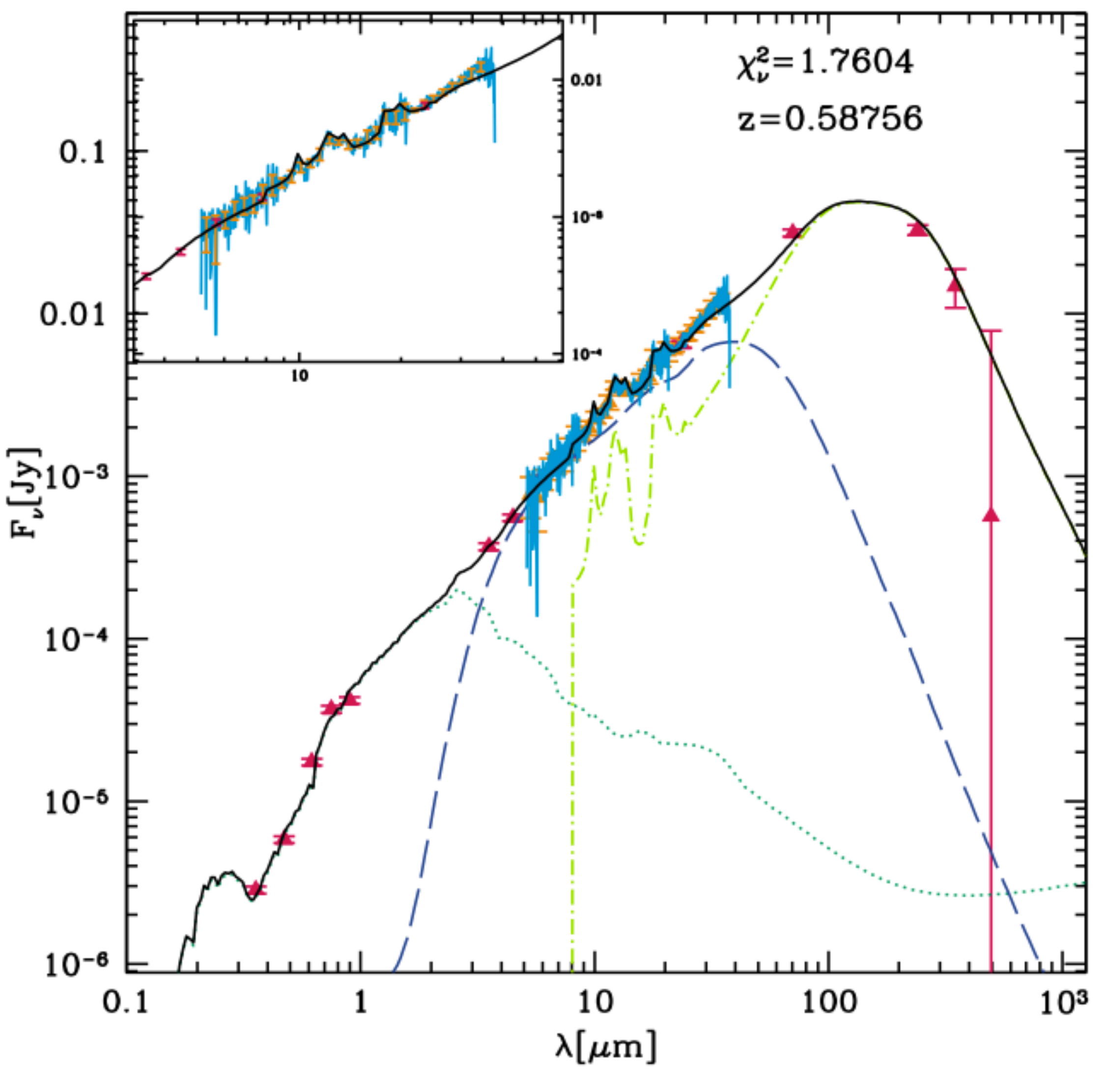}\hspace{1cm}
\includegraphics[width=6cm]{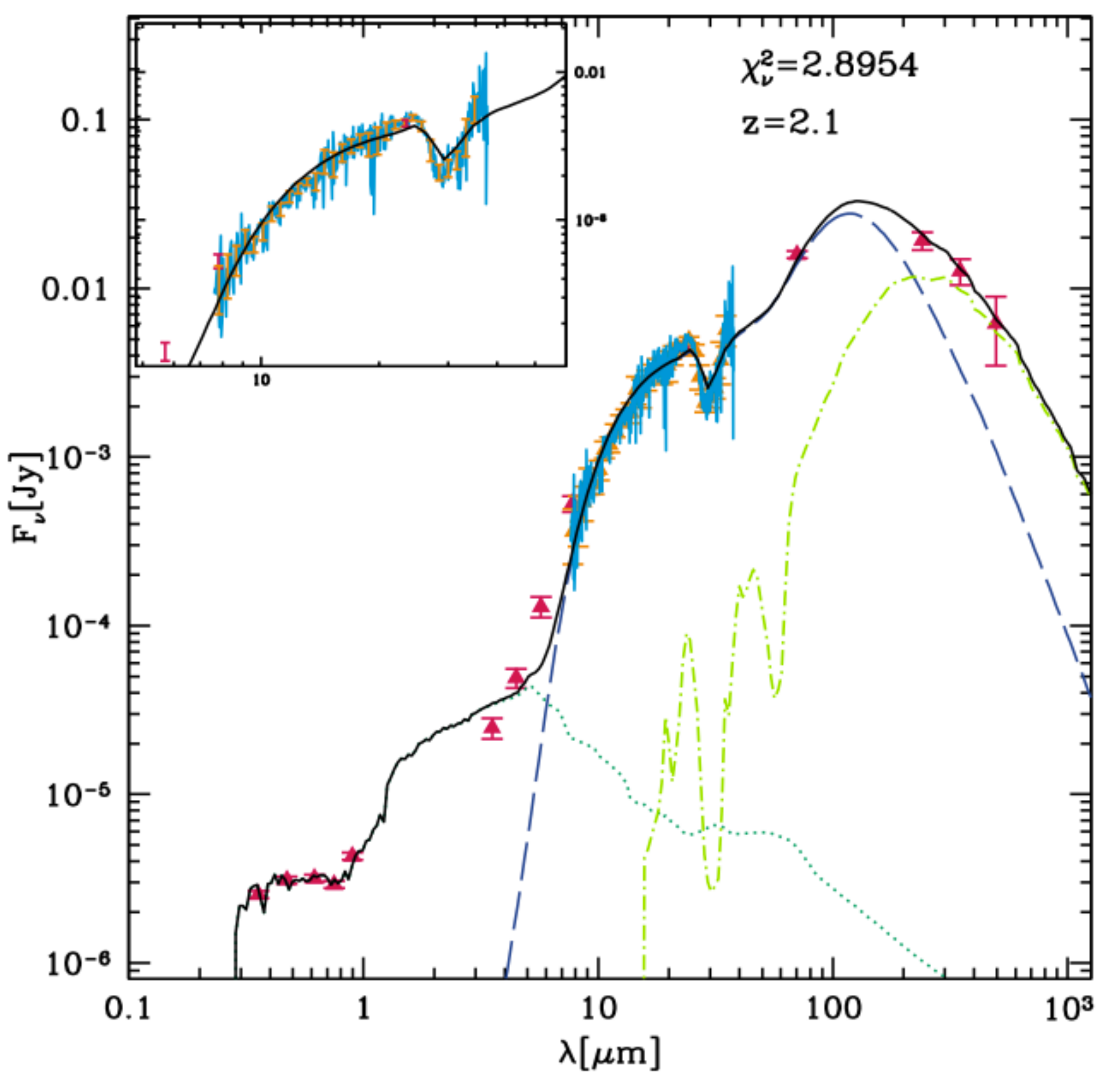}
\caption{Example of a best fit to a low (top) and a high (bottom) redshift object. The IRS spectrum (light blue; see also the insert figure) and photometric data (red symbols) are reproduced using the three emission components: SSPs (dotted dark green), AGN torus (dashed dark blue line) and starburst (dashed light green line), all together giving the total model emission (in black).}
\label{fig9}
\end{figure}

We have very recently made further improvements to a fully automatic fitting procedure, described for the first time in Hatziminaoglou et al. (2008) \cite{hatzimi08} and aimed to reproduce data from the optical/UV to the far-IR/submm. The procedure has been already applied in various published works \cite{hatzimi09,hatzimi10,vignali11,rodighiero11,pozzi12} and extensively tested with all the different options in the context of the {\itshape Herschel} Multi-tiered Extragalactic Survey (HerMES; PI S. Oliver). Three emission components (see also Fig. \ref{fig9}) are considered: 

\noindent - simple stellar population (SSPs) models, built using the Padova evolutionary tracks \cite{bertelli94}, account for the optical-UV emission; 

\noindent - AGN torus models, either smooth (F06) or clumpy (N08), to account for most of the mid-IR emission, due to hot dust heated by the presence of an AGN;

\noindent - empirical starburst galaxy templates to reproduce the bulk of the far-IR emission dure to the presence of cold dust.

A recently introduced improvement is the possibility to fit simultaneously mid-IR {\itshape Spitzer}/IRS spectra and broad band photometry. IRS spectra provide constrains on the torus models, especially when a silicate feature is present in emission or absorption, while {\itshape Herschel} photometry, especially in the long (SPIRE) wavelengths, allows us to constrain the peak and width of the cold dust emission. This can be seen in the illustrative examples of fit reported in Fig. \ref{fig9}. Furthermore, in order to compute the properties of the cold dust, such as the temperature and the mass we reproduced, in a second step, the far-IR emission with a grid of modified block body (with fixe value of $\beta=2$). 
This is a very powerful tool as it allows us to derive various physical properties for each of the components, such as the AGN accretion luminosity, the AGN and the starburst contributions to the total IR luminosity, the optical depth of the hot dust, the size and the mass of the torus, the stellar mass of the host galaxy and the star formation rate.

\subsection{Study of the AGN impact on a large sample}

We are now applying the full procedure to a large sample of $\sim$500 galaxies with multi-wavelngth photometry (from SDSS to {\itshape Herschel}) and IRS spectra available. The aim of this work is to study the impact of a possible presence of an AGN on the host galaxy's properties, analyzing the relative contribution of AGN to the total emission, the star formation rate in connection to the AGN and the mass of the hot (AGN) and cold (starburst) dust components. 

Preliminary results shows a consistency, despite a non-negligible scatter, between different estimates of the star formation rate (SFR) (e.g. from the IR luminosity, from PAH and from the fit to the optical data) and their independence from the AGN contribution to the total far-IR luminosity. Furthermore, no correlation has been found between the masses of the hot and cold dust and between the AGN accretion luminosity and the temperature of the cold dust (coming from the fit with the modified black body). 

\section{Discussion and Conclusions}

The conclusions we obtained from our results can be summarized as follows:

\noindent - the differences in the properties between F06 and N08 models arise mostly from the models assumptions (e.g. primary source, dust chemical composition) and not from the dust morphology with the consequence that the hot (AGN) dust properties, such as the size and the mass of the torus, strongly depend on the choice of the torus model;

\noindent - multi-wavelength SED fitting alone does not allow to distinguish between the two dust distributions;

\noindent -  the independence of the computed SFRs on the AGN contribution to the total far-IR luminosity indicates that the AGN component does not influence the star formation process;

\noindent - the absence of correlations found between the properties of the hot (AGN) and cold (SB) dust is an indication that the two dust heating mechanisms act on different scales. Indeed the central engine heats the dusty torus within few tens of pc at most, while the young stars in star forming regions, which are mainly responsible for cold dust heating, extends out to kpc scales.


%
%
\small  
%
\section*{Acknowledgments}   

%
This work makes use of the Nenkova et al. (2008a,b) models: \href{http://www.pa.uky.edu/clumpy/}{http://www.pa.uky.edu/clumpy/}. I thank R. Nikutta and M. Elitzur for providing detailed explanation on their models and on the calculation of the output parameters. I acknowledge support from ASI. SPIRE has been developed by a consortium of institutes led by Cardiff Univ. (UK) and including: Univ. Lethbridge (Canada); NAOC (China); CEA, LAM (France); IFSI, Univ. Padua (Italy); IAC (Spain); Stockholm Observatory (Sweden); Imperial College London, RAL, UCL-MSSL, UKATC, Univ. Sussex (UK); and Caltech, JPL, NHSC, Univ. Colorado (USA). This development has been supported by national funding agencies: CSA (Canada); NAOC (China); CEA, CNES, CNRS (France); ASI (Italy); MCINN (Spain); SNSB (Sweden); STFC, UKSA (UK); and NASA (USA). Funding for the creation and distribution of the SDSS Archive has been provided by the Alfred P. Sloan Foundation, the Participating Institutions, the National Aeronautics and Space Administration, the National Science Foundation, the U.S. Department of Energy, the Japanese Monbukagakusho, and the Max Planck Society. The SDSS Website is \href{http://www.sdss.org/}{http://www.sdss.org/}. This work makes use of TOPCAT (\href{http://www.star.bris.ac.uk/mbt/topcat/}{http://www.star.bris.ac.uk/mbt/topcat/}), developed by M. Taylor.

%

%
\end{document}